\documentclass[twoside]{photon2007}
\usepackage[latin1]{inputenc}
\usepackage[dvips]{graphicx,epsfig,color}
\usepackage{wrapfig,rotating}
\usepackage{amssymb,amsmath,array}

\usepackage[dvips]{color}

\newcommand{\EPEM}{\mbox{$e^+e^{-}$}}
\newcommand{\EMEM}{\mbox{$e^-e^-$}}
\newcommand{\GG}{\mbox{$\gamma\gamma$}}
\newcommand{\GE}{\mbox{$\gamma e$}}

\newcommand{\TEV}{\mbox{TeV}}
\newcommand{\GEV}{\mbox{GeV}}
\newcommand{\LGG}{\mbox{$L_{\gamma\gamma}$}}

\newcommand{\LEPEM}{\mbox{$L_{e^+e^-}$}}
\newcommand{\WGG}{\mbox{$W_{\gamma\gamma}$}}

\newcommand{\CM}{\mbox{cm}}
\newcommand{\MM}{\mbox{mm}}

\newcommand{\MKM}{\mbox{$\mu$m}}

\newcommand{\CMS}{\mbox{cm$^{-2}$s$^{-1}$}}

\newcommand{\ENX}{\mbox{$\epsilon_{nx}$}}
\newcommand{\ENY}{\mbox{$\epsilon_{ny}$}}

\newcommand{\be}{\begin{equation}}
\newcommand{\ee}{\end{equation}}
\newcommand{\bc}{\begin{center}}
\newcommand{\ec}{\end{center}}
\newcommand{\bi}{\begin{itemize}}
\newcommand{\ei}{\end{itemize}}
\newcommand{\ben}{\begin{enumerate}}
\newcommand{\een}{\end{enumerate}}

\begin{document}
\title{ Introduction to the Photon Collider} \author{V.I.~Telnov
% DO NOT MODIFY THE FOLLOWING '\vspace' ARGUMENT
\vspace{.3cm}\\
Budker Institute of Nuclear Physics, 630090 Novosibirsk, Russia \\
Novosibirsk State University, 630090 Novosibirsk, Russia 
}

\maketitle
\begin{abstract}
  The purpose of this Introduction, presented at 
PHOTON2007~\cite{TEL-PH2007}, is to provide an overview of the  
basic principles, possible parameters, some technical aspects
and the physics program of the photon collider and discuss its 
status within the ILC project.
\end{abstract}

\section{Introduction and politics}

There is a broad agreement within the world-wide particle physics 
community that the International Linear Collider (ILC) is the most
likely and desirable candidate for the next large HEP
project. If the LHC discovers new physics accessible to a 0.5--1.0 TeV
\EPEM\ linear collider, ILC
construction can start in 2012--2015. A photon (\GG, \GE) collider,
based on backward Compton scattering of laser light off high-energy 
electrons, is a natural extension of the linear
collider concept~\cite{GKST81,GKST8384,TEL90,TEL95}. It has been 
widely argued in the literature that addition of the photon collider
to the ILC design would nearly double the
ILC physics program, while the total cost of the ILC project would 
increase only by a few percent.

   The history of the photon collider is more than 25 years in the
making, as described in detail in my talk at
Photon2005~\cite{TELacta1}. The photon collider design was developed
and refined by enthusiasts in parallel with work on the \EPEM\ linear
collider; the LC and LCWS workshops and special Photon Collider
workshops provided a forum for discussions and brainstorming. About
20\% of papers on linear colliders are devoted to the photon collider
(mainly, the physics).

The basic principle of the photon collider is rather simple: it's just
Compton scattering. However, it took nearly two decades to gain 
understanding of the realistic photon-collider performance and to find 
solutions to the central technical challenges such as removal of 
disrupted beams from the
interaction region, mitigation of backgrounds, collision effects,
luminosity optimization, beam dump, requirements to the laser
parameters, the optical scheme and laser technologies that allow 
conversion of almost all electrons to photons, stabilization of beam
collisions, measurement of the luminosity, etc. We have now 
achieved understanding, at {\it a conceptual level}, of all 
critical issues in the design of the
photon collider and know how to solve them~\cite{TESLATDR,TELacta2}.

The photon collider was considered in the NLC~\cite{NLC}, JLC~\cite{JLC},
TESLA~\cite{TESLAcdr} conceptual design reports and in the TESLA
technical design report~\cite{TESLATDR}. Motivated to a great extent 
by the large cost of building a high-energy linear collider, in 2004 
the three regional projects were transformed into a single one, 
the ILC, which would be based on the superconducting technology 
developed by the TESLA collaboration. 

On one hand, unification of the three projects was a productive
development as it brought together the expertise of several
accelerator laboratories and financial resources of many countries. 
On the
other, as indicated by millenia of human experience in all kinds of
endeavors, presence of viable competition nearly always accelerates
the rate of progress, while absence of competition can sometimes be
detrimental.

It is interesting to consider the ILC schedule in the
context of the above observation. In the 1990s, and even at 
Snowmass 2001, NLC, JLC and TESLA were intended not merely
as complementary to the LHC, but also as LHC's contemporaries
and direct competitors. In fact, TESLA could have started operation 
in 2010! 

Since 2001, we have seen a dramatic revision of the ILC schedule: 2010
is now the year the ILC Engineering Design is due, with construction
beginning in 2012 and completing in 2019 being the most optimistic
projection.  U.S.~DOE officials expect a further delay beyond a
technology-driven timeline due to a long process of international
negotiations and the need to create the necessary legal framework for
the ILC; start of ILC operation in the late 2020s is now being floated.

Recognizing the danger of a long time gap between the scheduled end 
of Tevatron operation in 2009 and the start of ILC construction, 
Fermilab, which is considered as the most suitable site for the ILC
in the United States, has developed a back-up plan for the case of 
a prolonged ILC delay, dubbed \emph{Project X}: a 8 GeV superconducting 
proton linac for the study of neutrino physics and of help with promising 
future project (VLHC, muon collider, etc.) If \emph{Project X} is accepted,
a further delay in ILC construction is guaranteed. The question is, 
would the ILC, as it is envisioned today, still be relevant if it is 
built on a greatly delayed schedule?

Given the uncertain ability of the United States to host the ILC,
a number of alternatives will now be considered: Asia, Europe, Russia.  
Any construction decision would have to wait for results from the LHC. If
new physics below 0.5 TeV is found at LHC, a decision to start ILC
construction at one of the proposed sites can be made with little delay.

%%%%%

Let us now review the status of the photon collider within the ILC 
project. The ICFA Scope document on the ILC refers to \EPEM\ 
collisions at $2E=500$ GeV as ``the baseline'', while all other 
configurations (\EMEM, photon collider, Giga-$Z$,
operation at the $WW$ threshold, fixed-target, polarized $e^+$ beams) 
are considered ``options''. 
At the same time, the ICFA Scope document
dictates that the baseline ILC design must be made compatible with 
the future photon collider. Indeed, to make the photon collider 
possible in the second stage of the ILC project, it is extremely 
important from the very start to design the ILC to allow  
simple transitions between the \EPEM\ and \GG, \GE\ modes of operation
when they become available. 
 
Unfortunately, the compatibility with the photon collider was lost in 
preparation of the ILC Reference Design Report (RDR)~\cite{RDR}. Driven 
by a perceived need to reduce as much as possible the initial ILC cost, 
the RDR team considered only the basic \EPEM\ mode and was a bit too 
overzealous in cost-cutting. It made the unwise decision to propose a 
collider with a single IP and a 14 mrad crossing
angle, not compatible with the photon collider, which requires
a crossing angle of 25 mrad. 

While reducing the initial ILC cost by a few percent, the single-IP
solution with no space for the photon collider risked a great
escalation of the cost of upgrades, to no small part due to the need
for substantial additional excavation in the IP region half-way
through the ILC lifetime, which would be highly impractical and
perhaps technologically or politically impossible. It is obvious that
the total cost is minimal when all underground construction work is
done at once rather than in two or more stages, with their
considerable set-up costs and the disruption they would cause to ILC
operation.  Gravely concerned with the risks of the single-IP, 14 mrad
crossing-angle solution and its near incompatibility with the photon
collider, we strongly disagreed with this specific aspect of the RDR
and made an effort to change the situation.

And now, good news: shortly after PHOTON07, the GDE agreed that the
ILC Engineering Design should include the photon collider. At
IRENG07~\cite{ireng07} (September 2007), it was decided to correct the
layout of the interaction-region area in order to make it compatible
with \GG\ collisions. It would be still a single-IP configuration, but 
underground space will be reserved for an upgrade to the 25 mrad
crossing angle.  
So, the photon collider effort
is back on track after two years of uncertainties (and a struggle for
its very existence). Indeed, the years 2005--2007 were the most
difficult for the photon collider in its 25+ -year history. Now, we
can take a breath of fresh air, relax a bit, and continue
working towards making the photon collider a reality.

After this positive ``political information'', I turn to discussing the 
scientific problems of the photon collider.

\section{Principles and properties of the photon collider}

\subsection{The idea of the photon collider}

Two-photon physics had been talked about since 1930s, but as an active
research field is began in early 1970s, when production of \EPEM\
pairs was discovered in collisions of \emph{virtual} photons at an
\EPEM\ storage ring.  In the years that followed many interesting
two-photon reactions were studied, but the results could not compete
with the revolutionary discoveries made in \EPEM\ annihilation.  The
reason for this is that the luminosity and energy in virtual \GG\
collisions are small. Indeed, the number of equivalent photons
surrounding each electron is $dN_{\gamma}\sim 0.035 d\omega/\omega$,
and the corresponding \GG\ luminosity for $\WGG/2E_0>0.2$ is only
$\LGG \approx 4\times 10^{-3} L_{e^+e^-}$ and an order of
magnitude smaller for $\WGG/2E_0>0.5$.

The idea how to achieve much higher \GG\ luminosities was proposed by the
author of this paper at the First USSR workshop on the physics at the
linear collider VLEPP held in Novosibirsk in December 1980. 
Here it is in a nutshell: at linear colliders, beams are used only once,
which makes it possible to convert electrons to photons, and thus to
obtain collisions of \emph{real} photons. All that is needed is some sort of
a target a small distance from the interaction point (IP), where
the conversion would take place. For example, if one were to place a
target of $0.3 X_0$ thickness, the number of bremsstrahlung photons
would be greater than the number of virtual photons by one order of
magnitude, and the corresponding \GG\ luminosity would increase by two
orders of magnitude; however, this approach suffers from photo-nuclear
backgrounds. Laser light would make a much better target.

The method of production of high-energy photons by Compton scattering
of laser light off high-energy electrons was proposed in
1963~\cite{ARUTMilb} and soon afterwards was tested. However, the
conversion coefficient was very small, about $k=N_{\gamma}/N_e \sim
10^{-7}$. For the photon collider, we needed $k \sim 1$, seven orders
of magnitude more\,!

Soon after the 1980 VLEPP workshop, a group of \GG\ enthusiasts, I.~Ginzburg,
G.~Kotkin, V.~Serbo and V.~Telnov, considered the possibility of a
photon collider based on the laser conversion. We found that the
required flash energy is about 10 J.  Extrapolating the progress of
laser technologies into the next two decades with a high degree of optimism,
we came to the conclusion that a photon collider based on laser photon
conversion is not such a crazy idea after all, and in February 1981 
published the preprint, and then the paper~\cite{GKST81}. Somewhat latter, we
published two ``thick'' papers on this subject~\cite{GKST8384}.

The history of the photon collider and  from its origin to the present is
described in detail in \cite{TELacta1}.

\subsection{Basics of the photon collider}

Here, we briefly consider the main characteristics of backward Compton
scattering and the requirements on the lasers.

\vspace{-0.2cm}
\subsubsection{Kinematics and photon spectra \label{basics}}
In the conversion region, a laser photon of energy $\omega_0$
collides with a high-energy electron of energy $E_0$ at a small
collision angle $\alpha_0$ (almost head-on).  The energy of the
scattered photon $\omega$ depends on the photon scattering angle
$\vartheta$ with respect to the initial direction of the electron as
follows~\cite{GKST8384}: 
\begin{equation}
\omega = \frac{\omega_m}{1+(\vartheta/\vartheta_0)^2},
\label{kin1}
\end{equation} 
\vspace{-2mm}
\noindent where  
\vspace{-1mm}
$$\omega_m=\frac{x}{x+1}E_0, \;\;\;\;\vartheta_0= \frac{mc^2}{E_0}
\sqrt{x+1},$$
$$x=\frac{4E \omega_0 }{m^2c^4}\cos^2{\frac{\alpha_0}{2}} \simeq
 19\left[\frac{E_0}{\TEV}\right] \left[\frac{\mu
 \mbox{m}}{\lambda}\right],$$
\noindent $\omega_m$ being the maximum energy of scattered photons.
For example: $E_0 = 250$ GeV, $\omega_0 = 1.17$ eV ($\lambda=1.06$
\MKM) (for the most powerful solid-state lasers) $\Rightarrow$ $x=4.5$
and $\omega_m/E_0 = 0.82$.  Formulae for the Compton cross section can
be found elsewhere~\cite{GKST8384,TESLATDR}.

\begin{figure}[!tb]
\hspace{-0.8cm}\includegraphics[width=8.cm,height=5.8cm,angle=0]{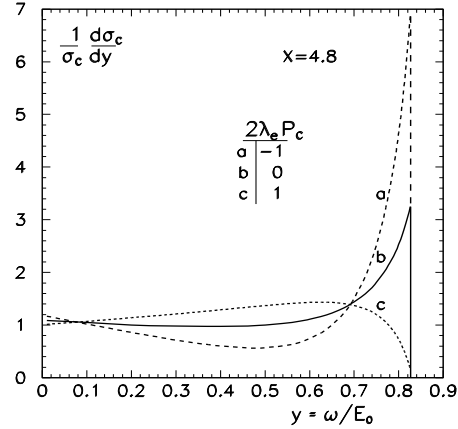}
\vspace{-8mm}
\caption{Spectrum of the Compton-scattered photons.}
\label{f1:fig4}
\vspace{-0.2cm}
\end{figure}

The energy spectrum of the scattered photons depends on the average
electron helicity $\lambda_{e}$ and that of the laser photons $P_c$.
The ``quality'' of the photon beam, i.e., the relative number of hard
photons, is improved when one uses beams with a negative value of
$\lambda_{e} P_c$.  The energy spectrum of the scattered photons for
$x=4.8$ is shown in Fig.~\ref{f1:fig4} for various helicities of the
electron and laser beams.
\begin{figure}[!tb]
\hspace{-0.5cm}\includegraphics[width=7.5cm,angle=0]{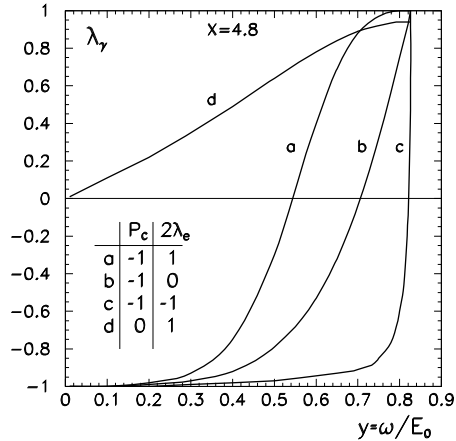}
\vspace{-8mm} \caption{Average helicity of the Compton-scattered photons.}
\label{f2:fig6}
\vspace{-0.3cm}
\end{figure}

With increasing $x$, the energy of the backscattered photons
increases, and the energy spectrum becomes narrower.  However, at
large values of $x$, photons may be lost due to creation of \EPEM\ 
pairs in collisions with laser photons~\cite{GKST8384,TEL90,TEL95}.
The threshold of this reaction is $\omega_m \omega_0 = m^2c^4$, which
corresponds to $x=2(1+\sqrt{2})\approx 4.83$.  One can work above this
threshold, but with a reduced luminosity; the luminosity loss factor
is in the 5--10 range for $x=$ 10--20. Therefore, $x\approx 4.8$ is the most
preferable value. The optimum wavelength of the laser photons
corresponding to $x=4.8$ is 
\begin{equation}
\lambda= 4.2 E_0 \;[\TEV]\;\; \MKM\,. 
\label{lamb}
\end{equation}
The mean helicity of backscattered photons at $x=4.8$ is shown in
Fig.~\ref{f2:fig6} for various helicities of the electron and laser
beams.  For $2 P_c \lambda_e = -1$ (the case of the peaked energy
spectrum), all photons in the high-energy peak have a high degree of
like-sign polarization.  A high degree of circular photon polarization
is essential for the study of many physics processes, for example, for
suppression of QED background in the study of the Higgs boson~\cite{TESLATDR}.

Linear polarization of backscattered photons is also possible at the
photon collider. The degree of the linear polarization at maximum
photon energy depends on parameter, it is $l_\gamma = 0.334,\ 0.6,\ 
0.8$ for $x= 4.8,\ 2,\ 1$, respectively~\cite{TESLATDR}.  The linear
polarization is important for the Higgs study .

\vspace{-0.2cm}

\subsubsection{Nonlinear effects in the conversion}
The electromagnetic field in the laser wave at the conversion region
is very strong, and so electrons can interact with several laser
photons simultaneously.  These nonlinear effects are characterized by
the parameter~(see~\cite{TESLATDR} and references therein) 
\begin{equation}
\xi^2 = \frac{e^2\overline{F^2}\hbar^2}{m^2c^2\omega_0^2} =\frac{2 n_{\gamma} r_e^2 \lambda}{\alpha}= 
\label{xi2}
\end{equation} \vspace{-3mm}
$$=0.36\left[\frac{P}{10^{18}\,\mbox{ W/cm$^2$}}\right]\left[\frac{\lambda}{\MKM}\right]^2,$$
\vspace{-0mm}
where $F$ is the r.m.s.\ strength of the electric (magnetic)
field in the laser wave and $n_{\gamma}$ is the density of laser photons.
At $\xi^2 \ll 1$, the electron scatters on one laser photon, while
at $\xi^2 \gg 1$ multi-photon scattering takes place.

The transverse motion of an electron through the electromagnetic wave
leads to the decrease of the maximum energy of the scattered photons:
$\omega_m/E_0 = x/(1+x+\xi^2)$.  At $x=4.8$, the value of
$\omega_m/E_0$ decreases by about 5\% for $\xi^2=0.3$.  For plots
demonstrating evolution of the Compton spectra as a function of
$\xi^2$ please refer to Refs.~\cite{Galynskii,TESLATDR}. With
increasing $\xi^2$, the Compton spectrum is shifted towards lower
energies and higher harmonics appear, and the \GG\ luminosity spectra
become broader. So, the value of $\xi^2\sim 0.3$ can be taken as the
limit for $x=4.8$; for smaller values of $x$ it should be even lower.

\vspace{-0.2cm}
\subsubsection{Laser flash energy}
While calculating the required flash energy, one must take into
account the diffractive divergence of the laser beam and to keep
the nonlinear parameter $\xi^2$ small. The r.m.s.\ radius of the laser beam
near the conversion region depends on the distance $z$ to the focus
(along the beam) as ~\cite{GKST8384}
\begin{equation} 
a_{\gamma}(z)=
a_{\gamma}(0)\sqrt{1+z^2/Z_R^2}\ , \;a_{\gamma}(0) \equiv
\sqrt{\frac{\lambda Z_R}{2\pi}},
\label{sLrz}
\end{equation}
where $Z_R$  is the Rayleigh length characterizing the length of
the focal region determined by focusing optics.
Neglecting multiple scattering, the dependence of the conversion
coefficient on the laser flash energy $A$ can be written as
\begin{equation} 
k  = N_{\gamma}/ N_e \sim 1-\exp (-A/A_0), \vspace*{-0.mm}
\label{kdef}
\end{equation}
where $A_0$ is the laser flash energy for which the thickness of the
laser target is equal to one Compton collision length. The value of
$A_0$ can be roughly estimated from the collision probability $p \sim
2 n_{\gamma}\sigma_{c}\ell = 1$, where $ n_\gamma \sim A_0/(\pi
\omega_0 a_{\gamma}^{2} \ell_\gamma )$, $\sigma_c$ is the Compton
cross section ($\sigma_c = 1.8\times10^{-25}$ cm$^2$ at $x=4.8$),
$\ell$ is the length of the region with a high photon density, which
is equal to $2Z_R=4\pi a_{\gamma}^2/\lambda$ at $Z_R \ll
\sigma_{L,z}\sim\sigma_z$ ($\sigma_z, \sigma_{L,z} $ are the r.m.s.\ 
lengths of the electron and laser bunches), and the factor 2 is due to
the relative velocity of electrons and laser photons. This gives, for
$x=4.8$,
\begin{equation} 
  A_0 \sim \frac{\pi\hbar c\sigma_z}{2\sigma_c} \sim 3 \sigma_z
  [\MM],\,\mathrm{J}. 
\label{A0estimate}
\end{equation}
Note that the required flash energy decreases when the Rayleigh length
is reduced to $\sigma_z$, but it hardly changes with further
decreasing $Z_R$. This happens because the density of photons grows
but the length of the high-density region decreases, and as result the
Compton scattering probability remains nearly constant. So, it is not
helpful to make the radius of the laser beam at the focus
smaller than $a_{\gamma}(0) \sim \sqrt{\lambda\sigma_z/2\pi}$, which
may be much larger than the transverse electron bunch size in the
conversion region. From (\ref{A0estimate}) one can see that the flash
energy $A_0$ is proportional to the electron bunch length, and for
$\sigma_z = 0.3$ mm (ILC) it is about 1 J.  The required laser power
is about half a terawatt.

Higher-precision calculations of the conversion probability in head-on
collision of a Gaussian laser beam with an electron beam can be found
elsewhere~\cite{GKST8384,TEL90,TEL95,NLC}; they are close to the above
estimate.

However, this is not a complete picture, since one should also take
into account the following effects:

$\bullet$ {\it Nonlinear effects in Compton scattering}. The photon
density is restricted by this effect. For shorter bunches, nonlinear
effects will determine the energy of the laser flash.

$\bullet$ {\it Collision angle}.  If the laser and electron beams
do not collide head-on (if the laser optics is outside the electron beam), the
required laser flash energy is larger by a factor of 2--2.5.

$\bullet$ {\it Transverse size of the electron beam}.  In the
crab-crossing scheme, the electron beam is tilted, which leads to an
effective transverse beam size comparable to the optimum laser spot size.
%\end{itemize}

\begin{figure}[!thb]
\vspace{-0.5cm}
\hspace{-0.5cm}\includegraphics[width=7cm,height=7.5cm]{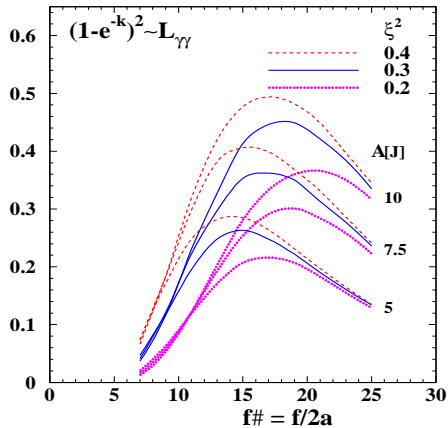}
\vspace{-1.1cm}
\caption{Dependence of  \LGG\ on the flash energy and $f_{\#}$
(flat-top laser beam) for several values of the parameter $\xi^2$. See comments in the text. }
\label{conversion}
\vspace{-0.1cm}
\end{figure}

Dependence of the \GG\ luminosity on the flash energy and
$f_{\#}=F/2R$ (flat-top laser beam) for several values of the
parameter $\xi^2$  is presented
in Fig.~\ref{conversion}~\cite{TEL-Snow2005,TELacta2}.  This simulation is
based on the formula for the field distribution near the laser focus
for flat-top laser beams. It was assumed that $\alpha_c = 25$ mrad and
the angle between the horizontal plain and the edge of the laser beam
is 17 mrad (the space required for disrupted beams and quads).
 At the optimum, $f_{\#} \sim 17$, or the
angular size of the laser system is about $\pm 0.5/f_{\#} \approx \pm
30$ mrad.  If the focusing mirror is situated outside the detector at
a distance of 15 m from the IP, it should have a diameter of about 1
m.  All other mirrors in the ring cavity can have smaller diameters;
about 20 cm seems sufficient from radiation-damage considerations.
This simulation as well as calculations done independently by the Zeuthen
group~\cite{Klemz2005}, show that with all effects taken into account, the
required flash energy for the photon collider at the ILC with
$2E_0=500$ GeV and for $\lambda=1.05$ \MKM\ is $A \approx 9$ J,
$\sigma_t \sim 1.5$ ps, $a_{\gamma}(0)=\sqrt{2}\sigma_{\gamma,x} \sim
10$ \MKM. The corresponding peak power is 2.5 TW.

The same laser with the 1 \MKM\ wavelength can be used up to an ILC
energy $2E_0 \sim 700$ GeV~\cite{TEL-Snow2005,TELacta2}. At higher
energies, the \GG\ luminosity decreases due to \EPEM\ pair creation in
the conversion region in collisions of the high-energy and laser
photons and due to the decrease of the Compton
cross section.  For the ILC energy $2E_0=1.0$ TeV, the reduction in the
luminosity due to these effects is about a factor of 2--3 compared to
the optimum case.  For $2E_0=0.7$--1 TeV, it is
desirable to have a wavelength of about 1.5--2 \MKM. The technical
feasibility of such a laser has not been studied yet.

\section{Interaction region issues}

\subsection{Collision scheme}

\begin{figure}[!bth]
\centering \vspace*{0.3cm}
\includegraphics[width=6.7cm,angle=0]{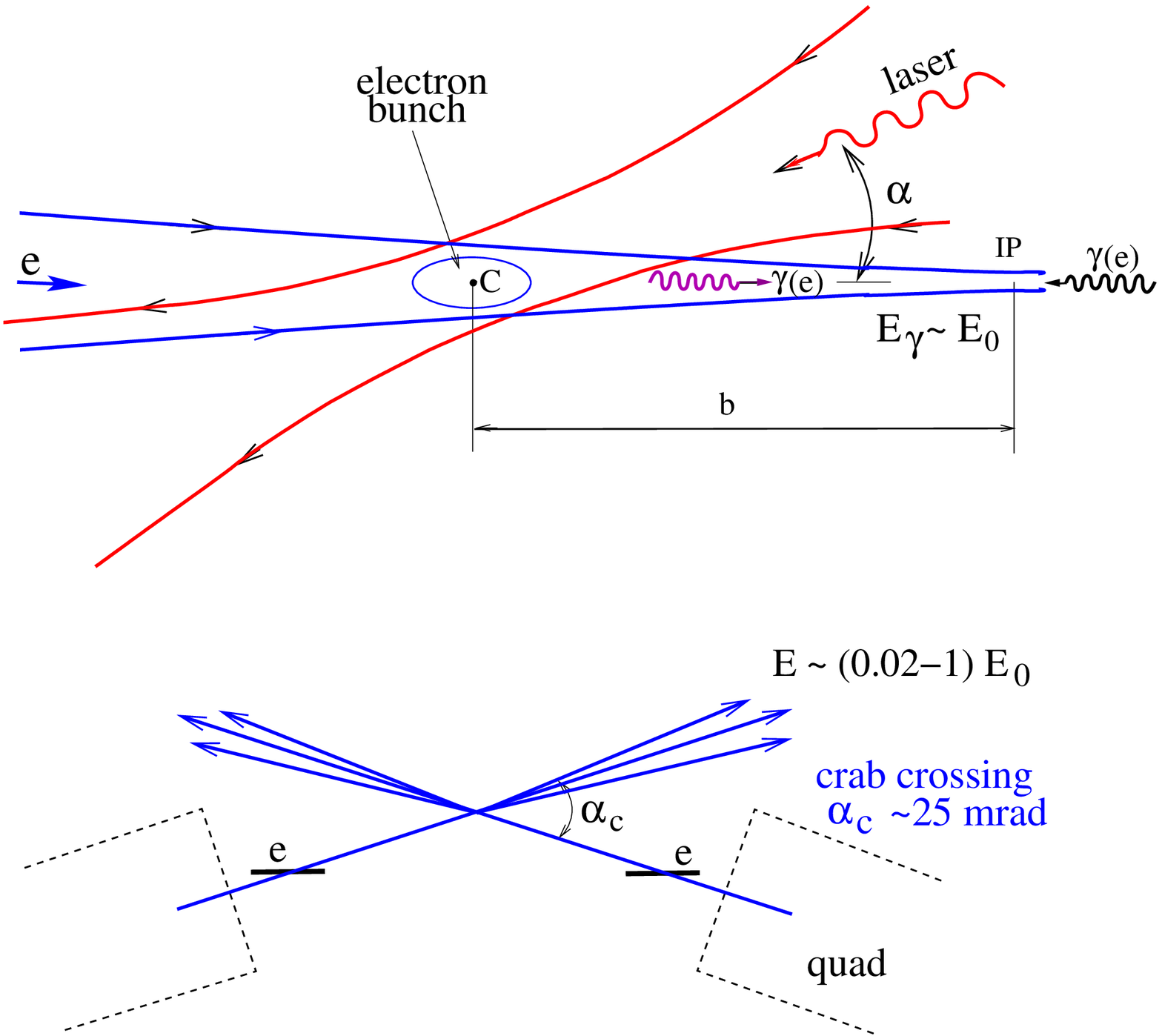}
 \vspace*{0.1cm}
\caption{Scheme of  \GG, \GE\ collider.}
\label{ggcol}
\vspace*{-0.0cm}
\end{figure}

The general scheme of the photon collider is shown in Fig.~\ref{ggcol}. The
laser light is focused on the electron beam in the conversion region
C, at a distance of $b$ cm from the interaction point IP; after
Compton scattering, the high-energy photons follow along the initial
electron trajectories with a small additional angular spread $\sim
1/\gamma$, i.e., they are in focus at the interaction point IP.
Thus obtained $\gamma$ beam collides downstream either with an oppositely 
directed electron beam or another $\gamma$ beam.

In the originally proposed scheme~\cite{GKST81,GKST8384}, 
following the conversion the electrons were swept away by a magnetic 
field $B\sim 1$ T.
The scheme with magnetic deflection of used beams allowed rather
clean \GG\ or \GE\ collisions to be produced. Taking $b \gg \gamma
a_e$, one can obtain a \GG\ luminosity spectrum with a width of
$\sim$10--15 \% (the ``monochromatization effect''~\cite{GKST8384}).
The optimum distance $b$ corresponds to the case when the size of the
photon beams at the IP due to Compton scattering is comparable to the
vertical (minimum) size of the electron beam: $b\sim \sigma_y
\gamma$. For the first linear collider projects, VLEPP and SLC, this
distance was about 10 cm, which was sufficient for magnetic deflection.

A year later, the vertical beam sizes in LC projects under consideration
were revised down to 3--5 nm.  For $\sigma_y=3$ nm, the optimum $b \sim
\gamma \sigma_y \sim 1.5$ mm for $2E_0=500$ GeV. This space is too
small to fit any kind of a magnet. Therefore, since 1991~\cite{TEL91}
we have been considering a scheme with no magnetic deflection. In
this case, there is a mixture of \GG, \GE\ and \EMEM\ collisions,
beamstrahlung photons give a very large contribution to the \GG\
luminosity at the low and intermediate invariant masses, the
backgrounds are larger, and the disruption angles are larger than in
the scheme with magnetic deflection (due to deflection of low-energy
particles in the field of the on-coming beam). However, there are
certain advantages: the scheme is simpler, and the luminosity is
larger. As for the backgrounds, they are larger but tolerable.

Note that even in the absence of deflecting magnets there is the beam-beam
deflection, which suppress the residual \EMEM\ luminosity. Also, at large
CP--IP distances and a non-zero crossing angle, the detector field
serves as the deflecting magnet and allows more-or-less clean
and quite monochromatic \GG, \GE\ collisions to be obtained with a reduced 
luminosity, which will be useful for QCD studies~\cite{TEL-mont}.

%\vspace{-0.2cm}
\subsection{The removal of beams}

 After crossing the conversion region, the electrons have a very broad
energy spectrum, $E=($0.02--1)\,$E_0$, and so the removal of such a
beam from the detector is far from trivial. In the scheme with
magnetic deflection, all charged particles travel in the horizontal
plane following the conversion.  At the IP, they get an additional
kick from the on-coming beam, also in the horizontal plane. This gave
us a hope that the beams can be removed through a horizontal slit in
the final quadrupoles. However it was not clear how to remove beams in
the scheme with no magnetic deflection.
\begin{figure}[!htb]
\vspace{-0.7cm}
\hspace{-0.3cm}\includegraphics[width=7.5cm]{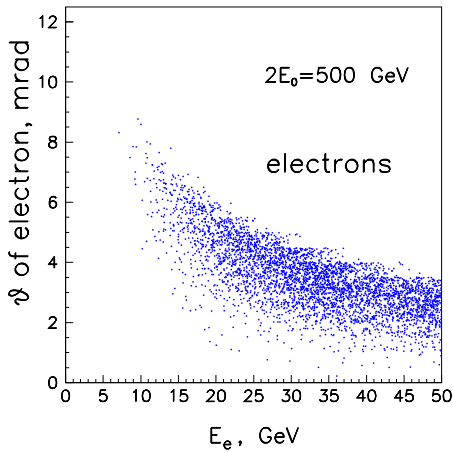}
 \vspace{-1.2cm}
\caption{Angles of disrupted electrons after Compton
    scattering and interaction with the opposing electron beam; $N=2\times
    10^{10}$, $\sigma_z=0.3$ mm.}  \hspace{0.5cm}
\label{e-angle}
\vspace{-0.5cm}
\end{figure}

In 1988, R.~Palmer suggested the crab-crossing scheme for \EPEM\ 
collisions at the NLC in order to suppress the multi-bunch
instabilities, Fig.~\ref{ggcol} (bottom).  In the
crab-crossing scheme, the beams are collided at a crossing angle
$\alpha_c$.  In order to preserve the luminosity, the beams are tilted
by a special cavity by the angle $\alpha_c/2$.  This scheme solves the
problem of beam removal at photon colliders~\cite{TEL90,TEL95}: the
disrupted beams just travel straight outside the quadrupoles.

In the scheme without magnetic deflection (which is now the primary
scheme), the low-energy particles get quite a large
deflection in the field of the opposing beam. The disrupted beams
have an angular spread of about $\pm$ 10 mrad (12 mrad with tails) after the
IP~\cite{TESLATDR,TEL-Snow2005,TELacta2}, see Fig.~\ref{e-angle}. The
disruption angle for low-energy particles is proportional to
$\sqrt{N/\sigma_z E}$~\cite{TEL90,TEL95} and depends very weakly on
the transverse beam sizes.  

The required crossing angle is determined by the disruption angle, the
outer radius of the final quadrupole (about 5
cm~\cite{TEL-Snow2005,TELacta2}), and the distance between the first
quad and the IP (about 4 m), which gives $\alpha_c = 12 +5/400 \approx
25$ mrad.

\subsection{The layout of the photon collider at the ILC}

For many years it was assumed that the future linear collider
would have two IPs, each equipped
with a detector, where the second IP, with a larger crossing angle, 
would be
optimized for the photon collider.  In the first few years, both
detectors would run in the \EPEM\ mode. Then, one of the IPs and 
the detector would be modified for operation in the \GG, \GE\ mode.

However, as was discussed in the Introduction, in the present ILC 
design~\cite{RDR} only one IP is planned, with a crossing angle of 14 mrad 
and two detectors in the pull-push configuration that could be swapped in
and out of the interaction region.
This is the minimum crossing angle that allows the outgoing beams in
\EPEM\ collisions to travel outside the quads. On the other hand, at 
the photon collider the crossing angle should be at least 25 mrad. 
This creates a problem.

A crossing angle greater than 30 mrad is not desirable because the
vertical beam size at the IP would increase due to synchrotron
radiation in the detector field. Yet at 25 mrad, the reduction of the
luminosity would be very small both for the \EPEM\ and \GG, \GE\ modes
of operation~\cite{TEL-LCWS05,TEL-Snow2005}. At first sight, it would
therefore seem quite reasonable to design the ILC with 25 mrad
crossing angle both for the \EPEM\ and the photon collider. Indeed,
why not? There are two arguments against it.

First of all, a smaller crossing angle is somewhat better for the study of
certain SUSY production processes in \EPEM\ collisions where detection of 
particles at small polar
angles is needed for suppression of the Standard Model backgrounds.
Then again, the difference between 14 and 25 mrad crossing angles is not 
that great.

The second, more serious contradiction between \EPEM\ and \GG\ has
to do with the difference in the requirements on the extraction
lines and beam dumps due to the very different properties
of electron and photon beams. 

In the \EPEM\ case, after collision the beams remain quite
monochromatic and there is a possibility to measure their properties
(the energy spectrum and polarization). Such an extraction line should
be quite long and equipped with many magnetic elements and
diagnostics.

At the photon collider, the situation is different:\\[-6mm]
\bi
\item  The disrupted beams at a photon collider consist of an equal mixture of
  electrons and photons (and some admixture of positrons);\\[-6mm]
\item Low-energy particles in the disrupted beams  have a large angular
 spread  and need exit pipes of a large diameter.\\[-6mm]
\item Following the Compton scattering, the photon beam
  is very narrow, with a power of about 10 MW. It cannot be dumped
  directly at any solid or liquid material. \\[-6mm]
\ei

\begin{figure}[!tbh]
 \centering
\includegraphics[width=6.8cm]{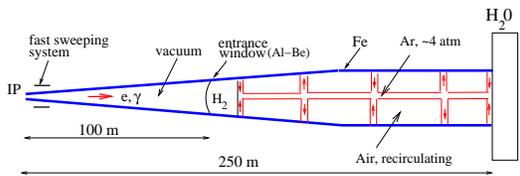}
\vspace{0.0cm}
\caption{An idea for the photon collider beam dump.  }
\vspace{-.3cm}
\label{beam-d}
\end{figure}

There exists an idea of a beam dump for the photon collider, as well
as some simulations~\cite{Telnov-lcws04,TEL-Snow2005,TELacta2}. In
short, it is a long tube, the first 100 m of which is vacuum, followed
by a 150 m long gas converter ending in a water-filled beam dump 
(Fig.~\ref{beam-d}). The diameter of the tube at the beam dump is about
1.5 m. In addition, there are fast sweeping magnet for electrons. Due
to a large beam width, no detailed diagnostics are possible, except perhaps
beam profile measurements.

\begin{figure}[!tbh]
\vspace{0.5cm}
\hspace{0.7cm}
\bc \includegraphics[width=6.8cm]{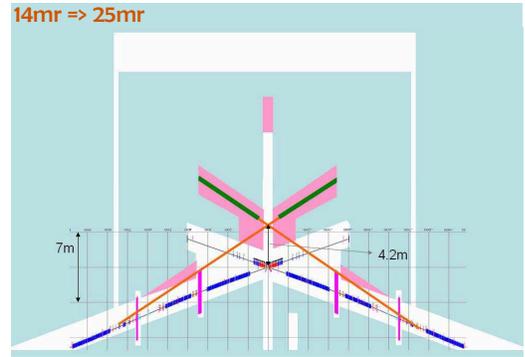} \ec
\vspace{0.2cm} \caption{The upgrade path from \EPEM\ to \GG\ (14
mrad to 25 mrad)} \label{f:seryi} \vspace{-0.05cm}
\end{figure}

So, the extraction lines and the beam dump for \EPEM\ and \GG\ are
 very different. Their replacement (transition to \GG\ and back after
 the energy upgrade) would be problematic due to induced
 radioactivity. Therefore it makes sense to have different crossing
 angles and separate extraction lines and beam dumps for \EPEM\ and
 \GG.  The suggestion of the ILC beam delivery group at LCWS06 was the
 following~\cite{Seryi-lcws06}.  For the transition from \EPEM\ to
 \GG, one has to move the detector and about 700 m of the up-stream
 beamline, Fig.~\ref{f:seryi}. The displacement of the detector required 
for the increase of the crab-crossing angle from
 14 to 25 mrad is about 4 m. The photon collider would also need 
an additional 250 m of tunnels for the beam dumps.

Such an upgrade would not be easy but in principle is acceptable, 
provided that all
excavation required for the photon collider is done at one time 
the initial ILC stage is built and the procedure of the
detector and beamline displacement is developed in advance.

In my opinion, expressed also at LCWS06~\cite{TEL-LCWS06-1}, the
presence of complicated diagnostic devices in the \EPEM\ extraction lines 
is not obligatory. Indeed, even without such special extraction lines we can
measure the beam energies and polarizations before collisions, many
relevant quantities can be measured during the collision 
(acollinearity angles,
distributions of the secondary \EPEM\ pairs, the beam deflection
angles); we can also measure the angular distributions and the charged and
neutral contents in the disrupted beams. All this allows the
reconstruction of the dynamics of beam collisions, with proper
corrections applied to the simulation. So, if one were to abandon the
idea of instrumented
extraction lines for \EPEM, the crossing angle, beam dumps and
beamlines for \EPEM\ and \GG\ could be the same, and the upgrade from
\EPEM\ to \GG\ would be much easier. 

The above suggestion was met by the advocates of the special, instrumented 
\EPEM\ extraction lines ``without enthusiasm''. For the sake of
a consensus, I accepted~\cite{TEL-LCWS06-1} the upgrade path from 14
to 25 mrad proposed by the ILC beam delivery group.

The \GG\ collider and other ``options'' are discussed in the 
Physics and Detector volumes of the ILC RDR. On the other hand, the 
Accelerator volume of the ILC RDR considered only the ``baseline''
configuration, making no mention of the upgrades and focusing on a
considerable reduction of the initial ILC cost, of which the \GG\ 
collider is only a small part. The intent was to reduce the cost without 
affecting the overall ILC physics program, and it was thought that the
extra excavation required for the photon collider can be done
at a later time. As explained above, this is not so, and therefore 
the photon-collider community could not possibly have been in 
in agreement with this opinion~\cite{TEL2007,Gronberg2007}.

Fortunately, common sense has prevailed. The next step in the ILC GDE 
is the Engineering Design Report. At the ILC Interaction region engineering
design workshop, IRENG07~\cite{ireng07}, we once again formulated
the requirements to the ILC design imposed by the upgrade path to the photon
collider~\cite{TEL-ireng07}, and they were accepted. The GDE team
agreed that the baseline ILC configuration should be modified in order
to make it compatible with the photon collider option, and all underground 
excavation work the photon collider would require should be done from the 
very beginning.

\subsection{The luminosity}
In \EPEM\ collisions, the maximum achievable luminosity is determined
by beamstrahlung and beam instabilities.  At photon colliders, the
only effect that restricts the \GG\ luminosity is the conversion of
the high-energy photons into \EPEM\ pairs in the field of the opposing
beam, that is, coherent pair creation~\cite{ChenTel,TEL90}. The
threshold for this effect $\kappa =(E_{\gamma}/mc^2)(B/B_0) \sim 1$,
where $B_0=\alpha e/r_e^2=4.4\times 10^{13}$ Gauss is the Schwinger
field and $B$ is the beam field. For \GE\ collisions, the luminosity
is determined by beamstahlung, coherent pair creation and the beam
displacement during the collision.

It is interesting to note that at the center-of-mass energies below 
0.5--1 TeV and for electron beams that are not too short, 
coherent pair creation is suppressed due to
the broadening and displacement of the electron beams during the
collision~\cite{TELSH-TSB2}: the beam field becomes
lower than the threshold for \EPEM\ production. So, one can
even use infinitely narrow electron beams.

All these processes, and a few
others, were included in simulation codes for beam collisions at linear
colliders. Results, presented below, were obtained by the
code~\cite{TEL95}, which was used for optimization of the photon
colliders at NLC~\cite{NLC} and TESLA~\cite{TESLAcdr,TESLATDR}.

\begin{figure}[!tbh]
\vspace{0.2cm}
\hspace*{-0.5cm}\includegraphics[width=8cm,angle=0]{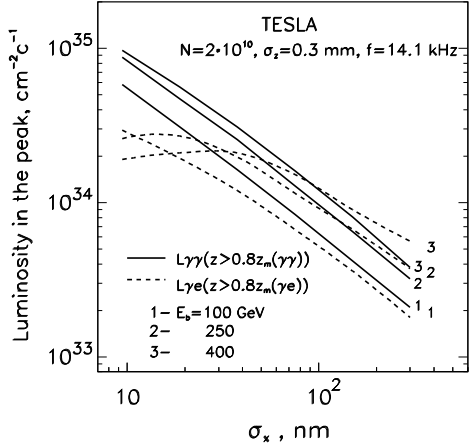}
\vspace{-0.5cm}
\caption{Dependence of \GG\ and \GE\ luminosities in
    the high energy peak on the horizontal beam size for TESLA-ILC
    at various energies.} 
\label{lumin}
\vspace{-0.4cm}
\end{figure}
Simulated values of \GG\ and \GE\ luminosities (in the high energy
peak) for TESLA (and, similarly, for ILC) are shown in
Fig.~\ref{lumin}~\cite{Tfrei,TEL2001,TESLATDR}. This figure shows how
the luminosity depends on the horizontal beam size (the vertical size
is much smaller). One can see that all \GG\ luminosity curves follow
their natural behavior: $L \propto 1/\sigma_x$. Note that for \EPEM,
the minimum horizontal beam size restricted by beamstrahlung is about
500 nm, while the photon collider can work even with $\sigma_x \sim
10$ nm at $2E_0=500$ GeV, delivering a luminosity that is several
times higher than that in \EPEM\ collisions! In fact, the \GG\
luminosity is simply proportional to the {\it geometric} \EMEM\
luminosity.

Unfortunately, the beam emittances in the damping-ring designs
currently under consideration cannot achieve beam sizes that are smaller
than $\sigma_x \sim$ 250 nm and $\sigma_y \sim 5$ nm, though a
reduction of $\sigma_x$ by a factor of two seems possible.  In
principle, one can use electron beams directly from low-emittance
photo-guns, avoiding the need for damping rings altogether, but at
present they offer a product of the transverse emittances that is
noticeably larger than can be obtained with damping rings (note: the
beams should be polarized).

To further reduce the beam emittances downstream of the damping rings
or photo-guns, one can use the method of laser cooling of the electron
beams~\cite{TELlasv,Tel-nano2002}. This method opens the way to
emittances that are much lower than those obtainable at damping
rings---however, this method requires a laser system that is much more
powerful than the one needed to achieve the $e \to \gamma$ conversion.
So, laser cooling of electron beams at linear colliders is a
technology for use at \GG\ factories in the distant future.

There is an approximate general rule: the luminosity in the
high-energy part of the spectrum $\LGG \sim 0.1 L_{ \rm
geom}$~\cite{TESLATDR}, where $ L_{ \rm geom}=N^2 \nu \gamma / 4\pi
\sqrt{\ENX \ENY\ \beta_x \beta_y}$. In order to maximize the
luminosity, one needs the smallest beam emittances \ENX, \ENY\ and
beta-functions at the IP, approaching the bunch length. Compared to
the \EPEM\ case, where the minimum transverse beam sizes are
determined by beamstrahlung and beam instability, the photon collider
needs a smaller product of horizontal and vertical emittances and a
smaller horizontal beta-function.

The ``nominal'' ILC beam parameters are: $N= 2\times 10^{10}$,
$\sigma_z=0.3$ mm, $\nu= 14100$ Hz, $\ENX = 10^{-5}$ m, $\ENY = 4
\times 10^{-8}$ m.  Obtaining $\beta_y \sim \sigma_z=0.3$ mm is not a
problem, while the minimum value of the horizontal $\beta$-function
is restricted by chromo-geometric aberrations in the final-focus
system~\cite{TESLATDR}. For the above emittances, the limit on the
effective horizontal $\beta$-function is about 5
mm~\cite{TEL-Snow2005,Seryi-snow}.  The expected \GG\ luminosity
$\LGG(z> 0.8z_m) \sim 3.5 \times 10^{33}$ \CMS\ $\sim 0.17\,\LEPEM$
(here the nominal $\LEPEM = 2\times 10^{34}$ \CMS)~\cite{TEL-Snow2005}.

\begin{figure}[!tbh]
\vspace{-1.2cm}
\hspace{-0.9cm}\includegraphics[width=8.5cm]{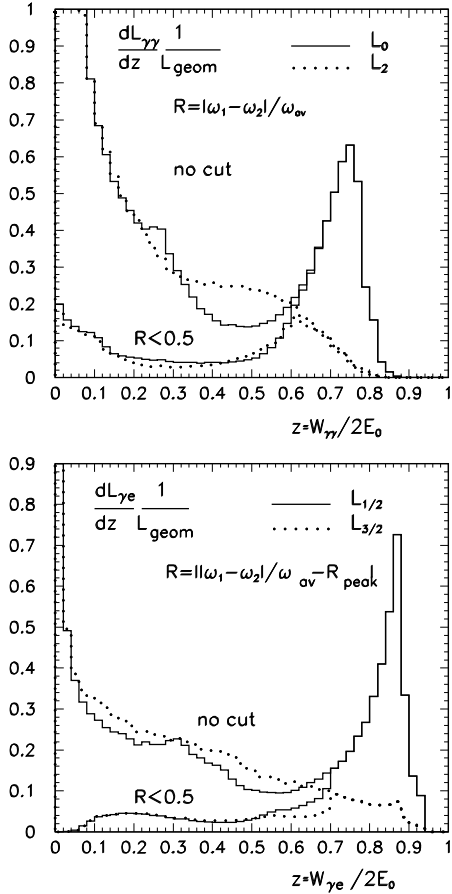}
\vspace{-1.7cm}
\caption{The \GG\ (upper) and \GE\ (bottom) luminosity spectra for
  typical TESLA (ILC) parameters at $2E_0=500$ GeV. Solid lines for
  $J_z$ of two colliding photons equal to 0, dotted lines for $J_z=2$
  (1/2 and 3/2, respectively, in the case of \GE\ collisions).  }
\label{lumspectra}
\vspace{-0.2cm}
\end{figure}

The typical \GG, \GE\ luminosity spectra for the TESLA-ILC(500)
parameters are shown in Fig.~\ref{lumspectra}~\cite{TESLATDR}. They
are decomposed to states with different spins $J_z$ of the colliding
particles. The total luminosity is the sum of the two spectra.  The
residual \EMEM\ luminosity (not shown) is one order of magnitude
smaller due to beam repulsion.  The luminosities with the cut on the
parameter $R$ are added just to show that the low-mass luminosities are
due to the very asymmetric collisions. One can see that \GG and \GE\
luminosities are comparable and these processes can be studied
simultaneously. 

However, it is much better to study \GE\ collision when only one of
the electron beams is converted to photons. In this case, one can
measure the \GE\ luminosity much more precisely~\cite{Pak} and with smaller
backgrounds. The problem of measuring the \GE\ luminosity spectra when
both beams are converted to photons (not completely) is due to the
uncertainty which direction the photon came from.

Taking into account the fact that cross sections for many interesting
processes are larger in \GG\ collisions than those in \EPEM\ by an
order of magnitude, the event rate in \GG\ collisions with nominal ILC
beams would be similar, or perhaps somewhat larger, than in \EPEM\ collisions.

However, it is a highly unsatisfying situation to have the \GG\ luminosity
limited by the beam emittances, an order of magnitude below its physics limit 
(determined by collision effects)! 
Being able to achieve the physics-limited \GG\ luminosity would open new 
physics possibilities, such as the
study of Higgs self-coupling in \GG\ collisions just above the $\GG\
\to hh$ threshold~\cite{Jikia}.  

It is an extremely interesting task to
search for a realistic technical solution for obtaining beams with smaller
emittances. There are a few good ideas, such as laser cooling, but
the first order of business 
should be trying to optimize the damping rings for the specific requirements
of achieving the highest possible luminosity at the photon collider. 
The importance of optimizing the damping ring design before the start of ILC
construction has been emphasized in~\cite{TEL-Snow2005,TELacta2,TEL-LCWS06-2},
the damping-ring experts and GDE management are aware of it, 
but, unfortunately, it appears that up to now the ILC damping-ring
design has been guided only by the baseline \EPEM\ considerations.

\subsubsection{Luminosity stabilization}
Beam collisions (luminosity) at linear colliders can be adjusted by
a feedback system that measures the beam-beam deflection using beam
position monitors (BPM) and corrects beam positions by fast kickers.
This method is considered  for \EPEM\ collisions and
is assumed for \GG\ as well~\cite{TESLATDR,TELacta2}.

There are some differences between the \EPEM\ and \GG\ cases.  In the
\EPEM\ case, at small vertical displacements the beams attract each
other and oscillate.  In the \GG\ case (\EMEM\ as well), the beams
repel each other; as a result, the deflection angle is larger and
almost independent of the initial displacement. There are also some
other differences connected with fluctuation of the conversion
efficiency. This problem and a stabilization algorithm were considered in
detail in Ref.~\cite{TELacta2}.

\subsubsection{Luminosity measurement}
The measurement of the luminosity at the photon collider is not an
easy task. The spectra are broad and one should measure the luminosity
and polarization as a function of energies $E_1, E_2$ of the colliding
particles~\cite{Pak}.  The luminosity spectrum and polarization can be
measured using various QED processes. These are  $\GG\to l^+l^-$
($l=e,\mu$)~\cite{TEL93,TESLATDR,Pak}, $\GG\to
l^+l^-\gamma$~\cite{Pak,Makarenko}  for \GG\ collisions and  $\GE\to\GE$ and
$\GE\to\;e^-\EPEM$ for \GE\ collisions~\cite{Pak}. Some other SM
processes could be useful as well.

\section{The laser and optics}

The photon collider at ILC(500) requires a laser system with the
following parameters: flash energy $A \sim
10$ J, $\sigma_t \sim 1.5$ ps, $\lambda \sim 1$ \MKM, and the following ILC
pulse structure: 3000 bunches within a 1 ms train and 5 Hz repetition
rate for the trains, the total collision rate being 15 kHz.  

In 1981, when the photon collider was proposed, the short-pulse Terawatt
lasers required for by a photon collider were just a dream.  A
breakthrough in laser technologies, the invention of the chirped pulse
amplification (CPA) technique~\cite{STRIC}, occurred very soon, in
1985.   The main problem in obtaining short pulses was
the limitation of the peak power imposed by the nonlinear refractive
index of the medium. This limit on intensity is about 1 GW/$\CM^2$;
the CPA technique successfully overcame it. 

The principle of CPA is as follows. A short, $\sim$ 100 fs low-energy
pulse is generated in an oscillator.  Then, this pulse is stretched by
a factor of $10^4$ by a pair of gratings, which introduces a delay
that is proportional to the frequency. This several-nanosecond-long
pulse is amplified, and then compressed by another pair of gratings
into a pulse of the initial (or somewhat longer) duration.  As
nonlinear effects are practically absent in the stretched pulses, the
laser pulses obtained with the CPA technique have a quality close to
the diffraction limit. This technique now allows the production of not
merely TW, but even PW laser pulses, and in several years the Exawatt
level will be reached, see the graph of laser power vs time in
\cite{TELacta1}.

The next, very serious problem was the laser repetition rate. The
pumping efficiency of traditional flash lamps is very low; the energy
is spent mainly on heating the laser medium. In addition, the
lifetime of flash lamps is too short, less than $10^6$ shots.
Semiconductor diode lasers solved these problems.  The efficiency of
diode laser pumping is very high, and heating of the laser medium is
low. The lifetime of the diodes is sufficient for the photon collider.

In addition to the average repetition rate, the time structure is of
great importance. The average power required of each of the two lasers
for the photon collider at the ILC is 10 J $\times$ 15000 Hz $\sim$
150 kW; however, the power within the 1 msec train is 10 J $\times
3000/0.001 \sim 30$ MW!  The cost of diodes is about ${\cal O }(1\$)
$/W, the pumping efficiency about 25\%, so the cost of just the diodes
would be at least ${\cal O }$\$100M), and the size of the facility
would be very large.

Fortunately, there is a solution. A 10 J laser bunch contains about
$10^{20}$ laser photons, only about $10^{11}$ of which are knocked out
in a collision with the electron bunch. So, it is natural to use the
same laser bunch multiple times. There are at least two ways to
achieve this: an optical storage ring and an external optical cavity.

In the first approach, the laser pulse is captured into a storage ring
using thin-film polarizers and Pockels
cells~\cite{NLC,TEL2001,TESLATDR}. However, due to the nonlinear
effects that exist at such powers, it is very problematic to use
Pockels cells or any other materials inside such an optical storage
ring.

\begin{figure}[!tbh]
\vspace{-0.0cm}
%\centering
\hspace{0cm}\includegraphics[width=7.2cm]{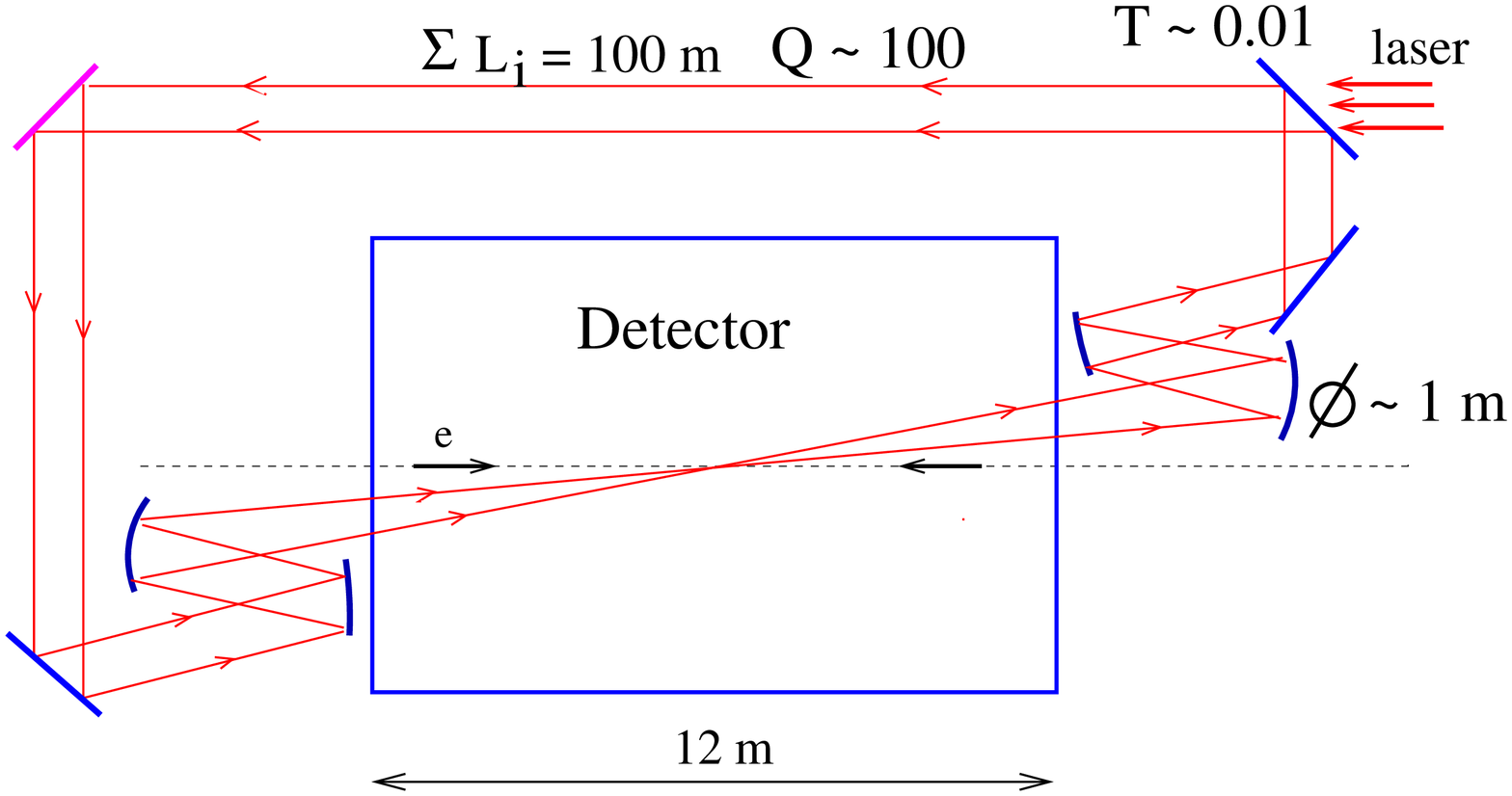}
%\vspace{-0.0cm}
\caption{External optical ring cavity for the photon collider}
\label{cavity}
%\vspace{-0.0cm}
\end{figure}

Another, more attractive approach, is an ``external'' optical cavity
that is pumped by a laser via a semi-transparent
mirror~\cite{Hiroshima1999,Tfrei,e-e-99,TEL2001,TESLATDR,Will2001,Klemz2005}.
One can create inside such a cavity a light pulse with an intensity
that is by a factor of $Q$ (the quality factor of the cavity) greater
than the incoming laser power. The value of $Q$ achievable at such
powers is 100--200. This means reduction of the required laser power
by a factor of hundred: for obtaining 10 J in the conversion region
one can use a laser with 0.1 J laser flash and pulse structure similar
to the ILC.  The optical-cavity principle is illustrated in
Fig.~\ref{cavity}. The cavity should also include adaptive mirrors and
other elements for diagnostic and adjustment.

Actually, such external cavities were known long ago in optical
laboratories, were used in a FEL experiments, in the
gravitational-wave experiment LIGO, etc., but were practically
unknown to HEP community: I arrived at this idea in
1999~\cite{Hiroshima1999,Tfrei,e-e-99} from the first principles, and
only later found that this technique already exists. 
Only then did I finally begin to believe in the technical feasibility 
of the photon collider
with the TESLA--ILC pulse structure and started to push it
vigorously~\cite{TEL2001,GG2000,TESLATDR}. Now, optical cavities are 
the baseline approach for the laser system at the ILC.

Note that the external optical cavity idea has proven to be a highly useful 
technique for HEP. It is now used for beam diagnostics (``laser wire''), 
for production of
polarized positrons for linear colliders (see Zomer's talk at this
conference~\cite{Zomer}) and can be used for laser cooling of electron
beams~\cite{TELlasv,Tel-nano2002}. 

Advancements in laser technologies are mostly driven by a handful of large,
well-funded programs, such as inertial-confinement fusion.  
These technologies are: the chirped-pulse
technique, diode pumping, laser materials with high
thermal conductivity, adaptive optics (deformable mirrors), disk
amplifiers with gas (helium) cooling, large Pockels cells, polarizers,
high-power and high-reflectivity multi-layer dielectric mirrors;
anti-reflection coatings, etc.  Now, practically all laser
technologies and components required for a photon collider are in
existence; nevertheless, the construction of such a state-of-the-art
laser system would not be an easy task.

\section{Physics}
 Although the \GG\ luminosity in the high-energy part of the spectrum
will be lower than in \EPEM\ by a factor of 3--5, the cross sections
in \GG\ collisions are typically greater by a factor of
5--10~\cite{TESLATDR}, so the number of ``interesting'' events would
surpass that in \EPEM\ collisions. Moreover, a further increase of the
achievable \GG\ luminosity by up to one order of magnitude 
may be possible.

Since the photon couples directly to all fundamental charged
particles---leptons, quarks, $W$'s, supersymmetric particles,
etc.---the photon collider provides a possibility to test every aspect
of the Standard Model, and beyond. Besides, photons can couple to
neutral particles (gluons, $Z$'s, Higgs bosons, etc.) through
higher-order diagrams. 

Many theorists took part in the development of the physics program for
the photon collider; the total number of publications has surpassed
the 1000 mark.

The physics program at the photon collider would be very rich and
complement in an essential way the physics in \EPEM\ collisions under
any physics scenario. In \GG, \GE\ collisions, compared to \EPEM,
\vspace{-0.1cm} \bi
\item the energy is smaller only by 10--20\%; \\[-7mm]
\item the number of interesting events is similar or greater; \\[-7mm]
\item access to higher particle masses (single resonances in $H$, $A$,
etc., in \GG, heavy charged and light neutral (SUSY, etc.) in
  \GE);  \\[-7mm]
\item at some SUSY parameters, heavy $H/A$-bosons will be seen only in
\GG;\\[-7mm]
\item higher precisions for some phenomena; \\[-7mm]
\item different types of reactions;  \\[-7mm]
\item highly polarized photons.  \ei

Some list of gold-plated processes is presented in Table~\ref{processes}.
\begin{table}[!hbtp]

\vspace{3mm}
{\renewcommand{\arraystretch}{1.} \small
\hspace{-0.3cm}
%\begin{center}
\begin{tabular}{ l  r } 

$\quad$ {\bf Reaction} &  {\bf  Remarks \hspace*{.5cm}} \\[3mm]
$\GG\to h^0 \to b\bar{b}$ &\hspace*{-1cm} $M_{h^0}<160$ \GEV  \\
$\GG\to h^0 \to WW(WW^*)$    & \hspace*{-0.5cm} $140<M_{h^0}<190\,\GEV$ \\
$\GG\to h^0 \to ZZ(ZZ^*)$      & \hspace*{-1.cm} $180<M_{h^0}<350\,\GEV$ \\
$\GG\to h^0 \to \GG\ $ & \hspace*{-1cm} $M_{h^0}<150$ \GEV  \\
$\GG\to H,A \to b\bar{b} $  &  \hspace*{-1.1cm} {\it MSSM} heavy Higgs \\
$\GG\to H\tau^+\tau^- $  &  \hspace*{-1.1cm} $\tan{\beta}$ in SUSY sector \\
$\GG\to \tilde{f}\bar{\tilde{f}},\
\tilde{\chi}^+_i\tilde{\chi}^-_i,\ H^+H^-$ &\hspace*{-1.cm} SUSY particles \\ 
$\GG\to S[\tilde{t}\bar{\tilde{t}}]$ & \hspace*{-1. cm}
$\tilde{t}\bar{\tilde{t}}$ stoponium  \\
$\GE \to \tilde{e}^- \tilde{\chi}_1^0$ & \hspace*{-1.3cm} $M_{{\tilde{e}^-}} < 
0.9\times 2E_0 - M_{{\tilde{\chi}_1^0}}$ \\
$\GE \to \tilde{\nu}_e \tilde{\chi}_1^{\pm} \to \tilde{\nu}_e \tilde{\nu}_\mu \mu$ & 
\hspace*{-1.3cm} sneutrino production \\
$\GE \to \nu W^+ \to W^+W^-l^+ $ & \hspace*{-1cm} Majorana neutrino \\

$\GG\to W^+W^-$ & \hspace*{-1.5cm} anom. $W$ inter., extra dim. \\
$\GE^-\to W^-\nu_{e}$ & \hspace*{-1.3cm} anom. $W$ couplings \\
$\GG\to WW+WW(ZZ)$ & \hspace*{-1.cm} strong $WW$ scattering \\
$\GG\to t\bar{t}$ & \hspace*{-1.3cm} anom. $t$-quark interact. \\
$\GE^-\to \bar t b \nu_e$ & \hspace*{-1.3cm} anom. $W tb$ coupling \\
$\GG\to$ hadrons & \hspace*{-1.3cm} total \GG\ cross section \\
$\GE^-\to e^- X$ and $\nu_{e}X$ &  \hspace*{-1.3cm} struct. functions \\ 
$\gamma g\to q\bar{q},\ c\bar{c}$ & \hspace*{-1.3cm} gluon distr. in the photon \\
$\GG\to J/\psi\, J/\psi $ & \hspace*{-1.3cm} QCD Pomeron \\

\end{tabular}
%\end{center}
}
\label{processes}
\caption{Gold-plated processes at PLC}
\end{table}
 More about physics at \GG\ colliders can be found in reviews
  \cite{TESLATDR,Boos,Velasco,DeRoeck,Krawczyk,Brodsky,Zerwas,Monig},
  references therein, and many other papers.

  So, the physics reaches of a \GG, \GE\ and \EPEM\ colliders are
  comparable. The only advantage of \EPEM\ collisions is the narrower
  luminosity spectrum, the feature that is useful but not obligatory.
  Also, the hadronic background ($\GG\ \to hadrons$) at the photon
  collider would be several times greater.  In \EPEM\ collisions,
 the beams at
  the IP also contain many beamstrahlung and virtual photons that
  produce hadrons.
  
  The photon collider can be added to the linear \EPEM\ collider at a
  very small incremental cost. The laser system and modification of
  the IP and one of the detectors would add about 3--4\% to the total
  ILC cost.  Some decrease of the \EPEM\ running time is a negligible
  price to pay for the opportunity to look for new phenomena in other
  types of interactions.

\section{Summary}

The physics expected in the 0.1--1 TeV region is very exciting, and the
ILC is a unique machine for the study physics in this energy
region. However, it is a very expensive machine, and therefore it should
strive to achieve ultimate performance and get maximum results. 
Answers to the mysteries of the origin of mass and the nature of the 
dark matter in the Universe would be a triumph of the entire mankind and 
would give excitement to several generations; from this perspective, 
\$10B or even \$30B is a negligible price to pay for these breakthroughs  in 
human understanding of the Universe.

 There is a very good chance that a linear collider will be built somewhere
in the world, and then the photon collider will inevitably happen, and the
study of new phenomena in \EPEM, \EMEM, \GG, \GE\ will bring us
to a new level of understand the world we live in!

\begin{footnotesize}

\end{footnotesize}
\end{document}